\documentclass[a4paper,fleqn]{cas-dc}

\usepackage{amsthm}
\usepackage[numbers]{natbib}
\usepackage{subfigure}
\usepackage{lineno,hyperref}
\usepackage{graphicx,floatrow}
\usepackage{subfigure}
\usepackage{amsfonts}
\usepackage{amssymb}
\usepackage{amsmath}
\usepackage{algorithm}
\usepackage{algpseudocode}
\usepackage{makecell,rotating,multirow,diagbox}
\usepackage{epstopdf}
\usepackage{color}
\usepackage{textcomp}
\usepackage{bm}
\usepackage{threeparttable}
\usepackage{booktabs}
\usepackage{graphicx}
\usepackage{longtable}
\usepackage{booktabs}
\usepackage{supertabular}
\usepackage{pdfpages}
\usepackage{calc}
\usepackage{url}
\usepackage{tikz}
\usepackage{lettrine}
\usepackage{epstopdf}
\usepackage{graphicx} 
\usepackage{tabularx}
\usepackage{listings}
\usepackage{xcolor}

\usepackage{xurl}

\usepackage{listings}
\usepackage{xcolor}

\newtheorem{definition}{Definition}
\newtheorem{mytheorem}{Theorem}

\begin{document}
\let\WriteBookmarks\relax
\def\floatpagepagefraction{1}
\def\textpagefraction{.001}
\shorttitle{xRWA: A Cross-Chain Framework for Interoperability of Real-World Assets}
\shortauthors{Guo et~al.}

\title [mode = title]{xRWA: A Cross-Chain Framework for Interoperability of Real-World Assets} 

\author[1]{Yihao Guo}[orcid=0000-0003-3266-6002]
\ead{yihao.guo@polyu.edu.hk}
\author[2]{Haoming Zhu}[orcid=0009-0005-1484-3361]
\ead{202300130137@mail.sdu.edu.cn}
\author[2]{Minghui Xu}[orcid=0000-0003-3675-3461]
\ead{mhxu@sdu.edu.cn}
\cormark[1]
\author[2]{Xiuzhen Cheng}[orcid=0000-0001-5912-4647]
\ead{xzcheng@sdu.edu.cn}
\author[1]{Bin Xiao}[orcid=0000-0003-4223-8220]
\ead{b.xiao@polyu.edu.hk}
\cormark[1]

\cortext[1]{Corresponding author}


\address[1]{Department of Computing, The Hong Kong Polytechnic University}
\address[2]{School of Computer Science and Technology, Shandong University}


\begin{abstract}        
Real-World Assets (RWAs) serve as a bridge between traditional financial instruments and decentralized infrastructures. By representing assets such as bonds, commodities, and real estate on blockchains, RWAs can extend the scope of decentralized finance. Industry forecasts further indicate rapid growth in tokenized RWAs after 2025, underscoring their potential role in the evolution of digital financial markets. However, in the current multi-chain environment, RWAs face challenges such as repeated authentication across multiple chains and inefficiencies arising from multi-step settlement protocols. To address these issues, we present a cross-chain framework for RWAs that emphasizes identity management, authentication, and cross-chain interaction. The framework integrates Decentralized Identifiers and Verifiable Credentials with customized attributes to support decentralized identification, and incorporates an authentication protocol based on Simplified Payment Verification to avoid redundant verification across chains. Furthermore, this paper adopts a cross-chain channel that supports efficient RWA settlements, and we refine its design so that the channel does not need to be closed immediately after each settlement, thereby reducing on-chain cost. We implement the framework and evaluate its performance via simulations, which confirm its feasibility and demonstrate improvements in efficiency for RWAs in cross-chain settings. 
\end{abstract}

\maketitle

\section{Introduction}
Real-World Assets (RWAs) refer to the tokenization of off-chain assets such as government bonds, commodities, and real estate, enabling them to be represented and transacted on blockchain systems~\cite{chen2024exploring}. 
They are attracting increasing attention in Web3 finance. As of June 2025, the total market capitalization of global stablecoins is approximately \$250 billion, while the market size of RWA tokenization grew from \$8.6 billion at the beginning of 2025 to \$23 billion in the first half of the year, representing a 260\% increase. 
Looking ahead, Citibank estimates that tokenized RWAs may reach a scale of around \$4 trillion by 2030, indicating their potential role in the future development of digital financial markets~\cite{RWA-data1,RWA-data2}. 

\begin{figure}[!t]
\centering
\centerline{\includegraphics[width=\textwidth]{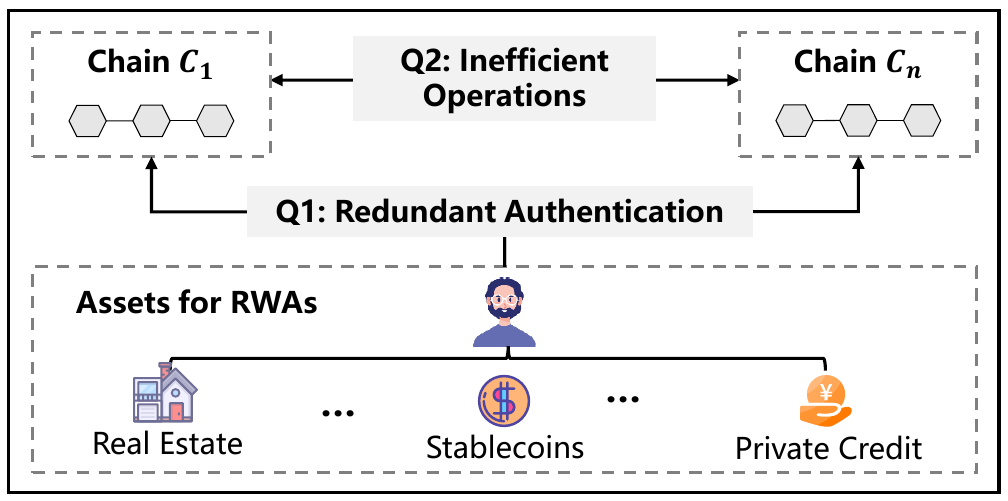}}
\caption{The multi-chain environment creates opportunities for the cross-chain circulation of RWAs. However, redundant authentication and inefficient cross-chain operations remain critical challenges.}  
\label{fig:problems}
\end{figure}

Blockchain~\cite{xu2024exploring,guo2022blockchain,gai2020blockchain} constitutes the key infrastructure for the tokenization and circulation of RWAs, providing transparency and immutability for asset issuance and transfer. In recent years, blockchain systems have expanded rapidly. By June 2025, CoinMarketCap reported more than ten thousand active cryptocurrency platforms~\cite{CoinMarketCap}, suggesting that RWAs can be deployed across an increasing number of heterogeneous blockchains. 
Although such distribution facilitates broader adoption~\cite{lao2020survey,guo2022blockchain,song2023survey}, the confinement of RWAs to single blockchains limits their financial utility, as assets cannot be traded across different chains. 
This situation is analogous to traditional financial markets where banks are unable to conduct transactions across borders, thereby restricting the mobility and efficiency of capital. Fortunately, cross-chain technologies~\cite{belchior2021survey,guo2023cross,guo2024zkCross}, regarded as a primary means of overcoming information silos between blockchains, may provide a feasible approach to enabling RWA interactions across heterogeneous systems. 
Nevertheless, existing research and practice on RWAs remain confined to single-chain contexts and have not systematically examined their cross-chain realization (shown in Figure~\ref{fig:problems}). 
To address this limitation, we propose a cross-chain RWA framework that aims to resolve the main challenges inherent in cross-chain processes. 

\noindent \textbf{Cross-Chain Redundant Authentication.}  
Compared with single-chain deployment, cross-chain scenarios require an RWA to undergo repeated authentications during inter-chain transfers. 
In a single-chain setting, an RWA only needs to be authenticated once within the local ledger, and its validity can then be consistently recognized across applications on that chain. 
In cross-chain environments, each blockchain maintains independent trust assumptions and verification rules. 
As a result, existing approaches typically require RWAs to undergo full authentication each time they are transferred on a different chain. 
Ideally, an RWA authenticated on one chain should be recognized by another through a proof of its prior authentication, thereby avoiding redundant procedures. 
Such a mechanism is particularly relevant when assets move from highly trusted chains, such as Ethereum mainnet, to less secure environments like its sidechains.  

\noindent \textbf{Inefficiency in Cross-Chain Operations.}  
Even within a single blockchain, operations involving RWAs already incur latency due to consensus.  
For instance, the two most prominent blockchain platforms, Bitcoin and Ethereum, support approximately 7 TPS and 15 TPS, which is significantly lower than traditional payment networks such as Visa, that can process several thousand TPS~\cite{visareport2023}.  
In cross-chain environments, this inefficiency is further amplified, since completing a cross-chain interaction typically requires multiple interdependent operations across heterogeneous ledgers.  
To ensure atomicity, decentralized protocols such as Hashed Timelock Contracts (HTLCs) are commonly employed; however, they generally involve four distinct on-chain operations, which substantially prolong completion time and increase operational costs.  
Cross-chain channels~\cite{guo2023cross,jia2023cross} attempt to mitigate this limitation by integrating HTLC-based mechanisms with off-chain channels to improve efficiency.  
However, in existing designs, each settlement requires closing the channel and opening a new one for subsequent interactions, which introduces additional costs and delays.  
Therefore, the central challenge is to adopt cross-chain channels that support efficient interoperability for RWAs. Moreover, it is necessary to design mechanisms that avoid repeated channel closures during settlement, thereby enabling lower on-chain costs. 

To fill this gap, we propose a cross-chain framework for RWAs to redefine the workflow of RWAs, with particular emphasis on identification, cross-chain authentication, and cross-chain interaction.  
To ensure decentralization throughout the process, we adopt Decentralized Identifiers (DIDs) and Verifiable Credentials (VCs), which are widely used in Web3, and extend them with customized attributes tailored to represent RWAs.  
Building on Simplified Payment Verification (SPV), we design a cross-chain authentication protocol that prevents redundant verifications of RWAs across multiple chains.  
Finally, we introduce a cross-chain channel scheme that enables the settlement of RWAs without requiring the channel to be closed after each settlement, thereby supporting efficient cross-chain interactions with lower costs. 

For convenience, we highlight our main contributions as follows: 
\begin{enumerate}
  \item We propose a cross-chain framework for RWAs that considers authentication and efficiency in cross-chain processes. To the best of our knowledge, this is the first work in this direction. 
  \item To address the problem of redundant authentication across multiple chains, we design a cross-chain authentication protocol that enables RWAs authenticated on one chain to be recognized by others. 
  \item 
  We employ a cross-chain channel to achieve more efficient RWA settlement. We further redesign the channel workflow to reduce operational overhead by eliminating the repeated opening and closing of channels. 
  \item We implement and conduct simulation experiments to demonstrate its feasibility and evaluate its performance. 
\end{enumerate}

The rest of the paper is organized as follows. 
In Section~\ref{sec:RW}, we review the most related work. Section~\ref{sec:pre} introduces the necessary preliminary knowledge. In Section~\ref{sec:main}, we present our framework and discuss its applications. Section~\ref{sec:per} reports simulation experiments that evaluate the performance of our scheme. Finally, Section~\ref{sec:con} concludes the paper with remarks and insights. 

\section{Related Works} \label{sec:RW} 
%
In this section, we review related work on RWAs from both industry and academia. 

\subsection{RWAs in Industry}
RWAs have already been actively deployed in various industry sectors. MakerDAO~\cite{MakerDAO} integrated tokenized U.S. Treasuries into its collateral framework, reinforcing the stability of decentralized finance. HSBC~\cite{rwa-gold} introduced a gold-backed RWA token for retail investors in Hong Kong. Securitize~\cite{Securitize} operates a regulated platform for issuing and trading RWAs, enabling compliant digital access to private companies. Brickken~\cite{BrickkenSeed} has tokenized RWAs worth over USD 250 million across 14 countries. Goldman Sachs and BNY Mellon~\cite{GoldmanBNY} partnered to tokenize shares of money market funds on  blockchains, aiming to improve accessibility and settlement efficiency. 

The RWA tokenization market has grown by more than 380\% from 2022 to 2025~\cite{RWA380}. These deployments now span heterogeneous blockchain infrastructures such as Ethereum, Polygon, Solana, and permissioned networks like Canton, resulting in RWA assets being distributed across isolated blockchain environments without shared standards for interoperability or cross-chain recognition.

\subsection{RWAs in Academia}
In academia, research on real-world assets has primarily focused on single-chain architectures and survey studies. 
Chen~{\em et al.}~\cite{chen2024exploring} analyzed the security of RWAs, highlighting risks in KYC/AML processes, oracle designs, and collateral structures. 
The work in \cite{koc2024escrow} investigated the connection between RWAs and escrow mechanisms, and discussed potential directions for future development. 
Zhao~{\em et al.}~\cite{zhao2025scalable} proposed a scalable framework that integrates Ethereum staking, Layer-2 rollups, and DAO governance for RWA management. 
Xia {\em et al.}~\cite{xia2025exploration} studied tokenization approaches for RWAs with the aim of enhancing liquidity and improving asset management practices. 
LEGO~\cite{ling2025sok} provided a quantitative methodology that maps historical failures in blockchain systems to current designs, including those involving RWAs. 

In academia, current research on real-world assets assumes a single underlying blockchain and lacks mechanisms for interoperability across heterogeneous chains. Existing studies do not address how RWAs deployed on different blockchain platforms can operate in multi-chain environments, resulting in fragmented deployments across multiple chains. 

\subsection{Summary and Motivations} 
Recent developments indicate that RWAs have already been deployed across multiple sectors of the financial industry, with initiatives ranging from tokenized government bonds and money market funds to gold and real estate. Meanwhile, academic research has largely concentrated on single-chain architectures and survey studies. This contrast reveals a critical gap: although RWAs are deployed on different blockchain systems, they remain isolated within their respective environments. Each implementation therefore operates in a silo, limiting the broader financial utility of tokenized assets. Motivated by this observation, we propose a cross-chain framework for RWAs that addresses the challenges of redundant authentication and low efficiency, and enables secure interoperability across multiple blockchains.

\section{Preliminaries} \label{sec:pre}
In this section, we present decentralized identifiers, verifiable credentials, simple payment verification, off-chain channels, and hashed timelock contracts, which serve as the building blocks of our work. 


\subsection{Decentralized Identifiers} \label{subsec:DID}
DIDs provide a decentralized identity mechanism for Web3, 
enabling users to prove ownership and control of their identities without relying on centralized authorities~\cite{w3c-did-core}. 
\begin{definition}[DID Operations]\label{Def:DID}
The operations of a DID are represented as a tuple of polynomial-time algorithms 
$\text{DID} \overset{\text{def}}{=} (\mathsf{KeyGen}, \mathsf{Create}, \mathsf{Resolve}, \mathsf{Update}, \mathsf{Deactivate})$: 
    \begin{itemize}
        \item $(pk,sk) \leftarrow \mathsf{KeyGen}(\mathsf{pp})$.  
        Given public parameters $\mathsf{pp}$, the algorithm generates a key pair $(pk,sk)$. 
        
        \item $(did,doc) \leftarrow \mathsf{Create}(pk,sk)$.  
        Given a key pair $(pk,sk)$, the algorithm outputs a new identifier $did$ and its associated document $doc$. 
        
        \item $doc \leftarrow \mathsf{Resolve}(did)$.  
        Given an identifier $did$, the algorithm returns the current document $doc$ linked to $did$. 
        
        \item $doc' \leftarrow \mathsf{Update}(did,doc)$.  
        Given an identifier $did$ and its document $doc$, the algorithm produces an updated document $doc'$, 
        reflecting revised verification methods or service endpoints under the controller’s authorization. 
        
        \item $\varnothing \leftarrow \mathsf{Deactivate}(did)$.  
        Given an identifier $did$, the algorithm deactivates it, making the identifier invalid and non-resolvable. 
    \end{itemize}
\end{definition}
DIDs have been widely adopted in Web3 ecosystems and are formally standardized by the World Wide Web Consortium (W3C)~\cite{mazzocca2025survey}.  
\subsection{Verifiable Credentials} \label{subsec:VC} 
VCs provide issuer-signed statements about a DID subject that can be verified by cryptographic proofs without trusting the presenter~\cite{w3c-vc-data-model-2}. 

\begin{definition}[VC Operations]\label{Def:VC}
Let the user be $\mathcal{U}$ with keys $(pk_{\mathcal{U}}, sk_{\mathcal{U}})$ and DID $did_{\mathcal{U}}$, 
the issuer be $\mathcal{I}$ with $(pk_{\mathcal{I}}, sk_{\mathcal{I}})$ and DID $did_{\mathcal{I}}$, 
and the verifier be $\mathcal{V}$. 
Let $\mathcal{R}^{\mathcal{I}}$ denote the issuer’s revocation list. 
The process is represented by the algorithm tuple 
$\Pi \overset{\text{def}}{=} (Req, \mathsf{Issue}, \mathsf{Prove}, \mathsf{Verify}, \mathsf{Revoke})$:
\begin{itemize}
  \item $req \leftarrow \mathsf{Req}(\{d_i\}_{\mathcal{U}}, sk_{\mathcal{U}})$. 
  Given data items $\{d_i\}$, $\mathcal{U}$ signs a request $req$ to obtain a credential. 
  \item $cred \leftarrow \mathsf{Issue}(req, sk_{\mathcal{I}})$. 
  Given $req$, $\mathcal{I}$ signs and returns a credential $cred$. 
  \item $\pi \leftarrow \mathsf{Prove}(cred, sk_{\mathcal{U}}, \phi)$. 
  Given $cred$ and a predicate $\phi$, $\mathcal{U}$ produces a proof $\pi$. 
  \item $\{0,1\} \leftarrow \mathsf{Verify}(\mathcal{R}^{\mathcal{I}}, cred, \phi, \pi, did_{\mathcal{I}})$. 
  $\mathcal{V}$ validates $\pi$, checks revocation against $\mathcal{R}^{\mathcal{I}}$, and confirms $did_{\mathcal{I}}$.

  \item $\mathcal{R}^{\mathcal{I}\,'} \leftarrow \mathsf{Revoke}(\mathcal{R}^{\mathcal{I}}, cred, sk_{\mathcal{I}})$. 
  $\mathcal{I}$ updates the status list to revoke or suspend $cred$.
\end{itemize}
\end{definition}
An important feature of VCs is selective disclosure, which allows a holder to reveal only specific attributes of a credential while keeping all other information hidden~\cite{xie2025slvc}.

\subsection{Simple Payment Verification}\label{subsec:SPV}
SPV enables a verifier to verify whether a transaction is included in the blockchain without downloading full blocks~\cite{xu2024exploring}. 

\begin{definition}[SPV]\label{Def:SPV}
The SPV process is represented by a tuple of polynomial-time algorithms 
$\text{SPV} \overset{\text{def}}{=} (\mathsf{Prove}, \mathsf{Verify})$.  
    \begin{itemize}
        \item $\pi \leftarrow \mathsf{Prove}(tx, B)$.  
        Given a transaction $tx$ and a block $B$ containing $tx$, the algorithm outputs a Merkle proof $\pi$, 
        which consists of a Merkle path $p$ authenticating $tx$ with respect to the Merkle root $\mu$ in the header of $B$. 
        
        \item $\{0,1\} \leftarrow \mathsf{Verify}(tx, \pi, \mu)$.  
        Given a transaction $tx$, a Merkle proof $\pi=(p,\mu)$, and a Merkle root $\mu$, the algorithm recomputes the authentication path using $p$ 
        and returns 1 if and only if $tx$ is correctly included in the block, and 0 otherwise. 
    \end{itemize}
\end{definition}

The Merkle path $p$ denotes the sequence of sibling hashes along the path from the transaction node to the root of the Merkle tree, 
and $\mu$ is the Merkle root contained in the corresponding block header.

\subsection{Off-Chain Channel} \label{subsec:oc}
An off-chain channel allows two parties to perform frequent transactions without recording each interaction on the blockchain, 
thereby reducing on-chain overhead while maintaining security guarantees through cryptographic commitments~\cite{ge2023accio}. 

\begin{definition}[Off-Chain Channel]\label{Def:OC}
The off-chain channel process can be represented by a tuple of polynomial-time algorithms 
$ \Pi \overset{\text{def}}{=} (\mathsf{Open}, \mathsf{Update}, \mathsf{Close})$. 
    \begin{itemize}
        \item $\mathsf{Open}$:  
        Both parties deposit assets into a contract on-chain, which locks funds as collateral for subsequent off-chain interactions. 
        
        \item $\mathsf{Update}$:  
        The parties exchange signed messages off-chain to update balances or states. 
        Each new state invalidates the previous one and is enforceable on-chain in case of disputes. 
        
        \item $\mathsf{Close}$:  
        When the channel is terminated, the latest signed state is submitted to the contract, 
        and the locked deposits are returned to the parties’ accounts according to this final state, 
        concluding the channel lifecycle.
    \end{itemize}
\end{definition}

Off-chain channels significantly improve scalability by shifting frequent interactions off-chain, 
while preserving security through cryptographic commitments and enforceability on-chain. 

\subsection{Hashed Timelock Contracts} \label{subsec:htlc}
HTLCs are conditional payment contracts widely used in cross-chain atomic swaps~\cite{guo2023cross}. 
They ensure that either both parties complete the transaction successfully or both are refunded, preventing unilateral loss. 

\begin{definition}[HTLC]\label{Def:HTLC}
HTLC consists of three polynomial-time algorithms  
$\mathsf{HTLC} \allowbreak \overset{\mathsf{def}}{=} (\mathsf{Lock}, \mathsf{Unlock}, \mathsf{Refund})$. 
It operates on two blockchains $\mathbb{B}_0$ and $\mathbb{B}_1$, where both parties deploy 
contracts with the same hash value and coordinated timeouts. 

\begin{itemize}
    \item \textbf{$\mathsf{Lock}$}:  
    The initiator $P_0$ samples a random preimage $\rho$ and computes its hash $h(\rho)$.  
    $P_0$ locks an asset on $\mathbb{B}_0$ in a contract that specifies $h(\rho)$ and a timeout $T_0$.  
    The counterparty $P_1$ then deploys a corresponding contract on $\mathbb{B}_1$ using the same hash value 
    and a shorter timeout $T_1 < T_0$.
    
    \item \textbf{$\mathsf{Unlock}$}:  
    To redeem the asset on $\mathbb{B}_1$, $P_0$ reveals $\rho$ to the contract on $\mathbb{B}_1$ before $T_1$.  
    Since all blockchain data are public, $P_1$ learns $\rho$ from this on-chain revelation and submits it 
    to the contract on $\mathbb{B}_0$ before $T_0$ to unlock the asset.  
    Both contracts verify that the submitted preimage satisfies $h(\rho)$ before releasing funds.
    
    \item \textbf{$\mathsf{Refund}$}:  
    If either party does not obtain $\rho$ before the corresponding timeout,  
    the contract on $\mathbb{B}_0$ (resp. $\mathbb{B}_1$) refunds the asset to $P_0$ (resp. $P_1$),  
    ensuring no loss in the event of failed coordination.
\end{itemize}
\end{definition}

HTLCs implement cross-chain payments by enforcing that assets are released only when the correct preimage is disclosed, and otherwise returned to the sender after the timeout, thereby ensuring atomicity.  

In the following section, we adopt these techniques to design our proposed scheme.

\section{xRWA} \label{sec:main}
This section begins with an overview of xRWA with its design goals, followed by a detailed description of the proposed scheme. 
\subsection{Overview}

\begin{figure}[!htbp]
\centering
\centerline{\includegraphics[width=\textwidth]{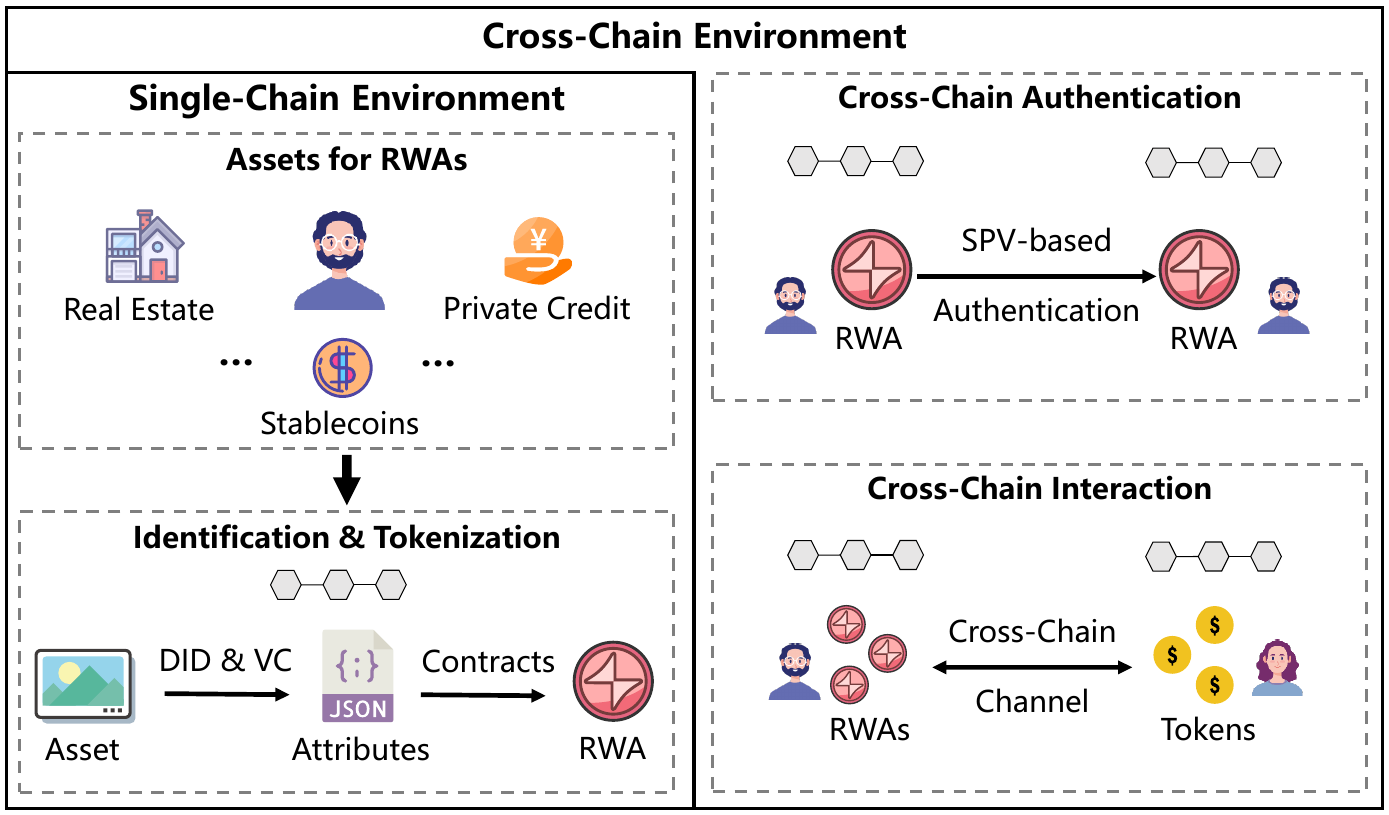}}
\caption{xRWA comprises three functional layers: (i) asset identification and tokenization, where DIDs and VCs are used to represent real-world assets on-chain; (ii) cross-chain authentication, which prevents repeated registration across heterogeneous blockchains; and (iii) cross-chain interaction through channel-based settlement, enabling efficient trading of RWAs.} 
\label{fig:overview}
\end{figure}

xRWA provides a framework for RWAs that supports secure identification, cross-chain authentication, and efficient interoperability across heterogeneous blockchains. Figure~\ref{fig:overview} shows the workflow, which consists of three layers: asset identification and tokenization, cross-chain authentication, and cross-chain interaction.   

The process begins with the digital representation of off-chain assets such as real estate, stablecoins, and private credit assets. Traditional approaches rely on centralized custodians or special-purpose vehicles that create contractual mappings between physical assets and their digital forms. xRWA adopts DIDs and VCs for decentralized identity management. By embedding customized attributes into VCs, the framework binds issuers and holders to asset records, ensuring legitimacy while reducing the reliance on intermediaries.  
After tokenization, assets may need to be recognized across multiple blockchains. Existing methods often require repeated verification on each chain. xRWA introduces a cross-chain authentication protocol based on SPV proofs. By linking DIDs and VCs with tokenized assets, and applying SPV validation, authentication performed on one chain can be reused across others.  
Cross-chain transfers also require efficient settlement. xRWA proposes a cross-chain channel that supports settlement without channel closures. This allows frequent interactions to occur off-chain, with only final states recorded on-chain, reducing on-chain costs. 


\subsection{Design Goals}
\label{subsec:rwa-security}

The security objective of xRWA is determined by two core guarantees that capture 
its correctness in multi-chain environments. 

\noindent{\bf Cross-chain authentication correctness.} 
A blockchain participating in xRWA may authenticate an RWA either (i) directly 
on its own ledger or (ii) by accepting an authentication result originating 
from another chain. For the second case, correctness requires that a chain $C_t$ 
accept such a cross-chain authentication claim only if it corresponds to a 
genuine and confirmed authentication record on the source chain $C_s$ according 
to the protocol rules. In particular, no PPT adversary should be able to 
fabricate cross-chain evidence so as to make $C_t$ accept an RWA based on a 
non-existent or invalid authentication record.

Let $\mathsf{AuthAcc}_{C_t}(a)$ denote the event that chain $C_t$ accepts 
asset $a$ via a \emph{cross-chain} authentication claim from $C_s$, and let 
$\mathsf{AuthRec}_{C_s}(a)$ denote the existence of a valid and confirmed 
authentication record for $a$ on $C_s$. Authentication correctness requires 
that for every PPT adversary $\mathcal{A}$,
\[
  \Pr\!\left[
    \mathsf{AuthAcc}_{C_t}(a)
    \;\wedge\;
    \neg\,\mathsf{AuthRec}_{C_s}(a)
  \right]
  \leq \mathsf{negl}(\lambda),
\]
meaning that a cross-chain claim should never be accepted unless it is 
consistent with an actual authentication on the source chain.

\noindent{\bf Cross-chain settlement atomicity.}
When two blockchains engage in an RWA settlement procedure, the resulting state
changes must be all or nothing. If settlement succeeds, both ledgers must apply 
the corresponding asset and payment updates. If settlement fails, both ledgers 
must retain their original states, and no honest party should end up with a 
partially executed outcome.

Let $\mathsf{Com}_1$ and $\mathsf{Com}_2$ denote the events that chains $C_1$ 
and $C_2$, respectively, apply the settlement updates for a given cross-chain 
interaction. Atomicity requires that for every PPT adversary $\mathcal{A}$,
\[
  \Pr\!\left[
    \mathsf{Com}_1
    \;\oplus\;
    \mathsf{Com}_2
  \right]
  \leq \mathsf{negl}(\lambda),
\]
where $\oplus$ denotes exclusive-or. Thus, it is infeasible for an adversary to 
cause exactly one chain to commit while the other does not, or to produce any 
partial or inconsistent settlement outcome.

\begin{definition}[RWA Security]
\label{def:rwa-security}
A cross-chain RWA protocol $\Pi$ is said to satisfy \emph{RWA security} if it 
ensures both (i) cross-chain authentication correctness and (ii) cross-chain 
settlement atomicity. Namely, for every PPT adversary $\mathcal{A}$, the 
probability that $\mathcal{A}$ causes a chain to accept a fabricated 
cross-chain authentication claim is negligible, and the probability that 
$\mathcal{A}$ produces an inconsistent or partially executed cross-chain 
settlement across two blockchains is negligible in the security parameter 
$\lambda$. 
\end{definition}

%
\subsection{Identification and Tokenization}


To realize the identification of RWAs, we construct $\mathsf{RWA\text{-}CC}$ (RWA Composite Credential) using DIDs and VCs. The credential follows a modular design with four components: $\mathsf{asset}$, $\mathsf{identity}$, $\mathsf{compliance}$, and $\mathsf{custody}$. Each component includes its own $\mathsf{sStatus}$ and $\mathsf{sProof}$ to support independent revocation and verification, while a top-level proof guarantees overall integrity. 

The $\mathsf{asset}$ component defines the core identification and tokenization of a real-world asset. 
The top-level $\mathsf{id}$ specifies the identifier of the credential instance itself. 
Within the $\mathsf{asset}$ object, the field $\mathsf{assetId}$ assigns a DID-based unique identifier to the asset. 
The field $\mathsf{assetType}$ indicates the broad category of the real-world asset, while $\mathsf{category}$ provides a finer classification within that type, for example distinguishing $\mathsf{Residential}$ from $\mathsf{Commercial}$ property. 
The $\mathsf{classDid}$ links the asset to a class definition maintained by an authoritative registry, enabling semantic grouping and interoperability across assets of the same class.  
A crucial element is $\mathsf{tBinding}$, which establishes the binding between the off-chain asset record and its tokenized representation on a blockchain. 
It records the token standard, the target chain, the contract address, and the token identifier in a unified structure.  

The $\mathsf{identity}$ component extends the description of an asset beyond its basic identifier. 
The field $\mathsf{schemaVersion}$ specifies the version of the identity schema, while $\mathsf{identitySchema}$ provides a reference URI that defines its structure. 
The $\mathsf{identifiers}$ array records registration information from official registries: each entry contains details such as the registration scheme, the identifier value, the jurisdiction, and the issuing authority. 
The following components may incorporate attributes specifically designed to uniquely identify various categories of RWAs, such as location, applicable standards, associated documentation, or other distinctive characteristics. These attributes record the intrinsic properties of the RWA and establish a one-to-one correspondence between the VC and the RWA entity.
The $\mathsf{compliance}$ component defines the regulatory framework governing an asset’s circulation, including its authorizing license, permitted jurisdictions, applicable restrictions, validity period, and the regulator’s identifier. Similarly, the $\mathsf{custody}$ component records safekeeping information: custodian identity, storage location, operational policy, audit frequency, and insurance reference. 

After an RWA is identified through $\mathsf{RWA\text{-}CC}$, the next step is to tokenize it on a blockchain. 
The process builds on the algorithms for DIDs and VCs introduced in Section~\ref{sec:pre}.  
The issuer $\mathcal{I}$ generates its decentralized identity using $\mathsf{DID.KeyGen}$ and obtains a DID $did_{\mathcal{I}}$. 
The asset $a$ is also assigned a DID $did_{a}$, which is included in the $\mathsf{asset}$ component of $\mathsf{RWA\text{-}CC}$.  
The issuer then executes $\mathsf{VC.Issue}$ to generate a credential $\mathsf{cred}$ that encapsulates the attributes of $a$, including $\mathsf{identity}$, $\mathsf{compliance}$, and $\mathsf{custody}$. 
This binds the off-chain asset to its verifiable description under the issuer’s DID.  
To make the asset tradable, the credential is linked to an on-chain token via $\mathsf{tBinding}$, which specifies the token standard (e.g., ERC-721), the blockchain, the contract address, and the identifier. 
This ensures that smart contracts interacting with the token can invoke $\mathsf{VC.Verify}$ against the issuer’s DID and status list to check its validity.  
Finally, when a holder $\mathcal{U}$ presents the token in applications, only the necessary attributes are disclosed through $\mathsf{VC.Prove}$, ensuring selective transparency while maintaining compliance guarantees.  
\begin{lstlisting}[
  basicstyle=\ttfamily\small,
  showstringspaces=false,
  breaklines=true,
  frame=single,
  backgroundcolor=\color{gray!7},
  rulecolor=\color{black!40},
  numbers=left, numberstyle=\tiny\color{gray}, stepnumber=1, numbersep=6pt,
  stringstyle=\color{green!50!black},
  literate=
   *{true}{{{\color{blue}true}}}{4}
    {false}{{{\color{blue}false}}}{5}
    {null}{{{\color{blue}null}}}{4}
    {0}{{{\color{blue}0}}}{1}
    {1}{{{\color{blue}1}}}{1}
    {2}{{{\color{blue}2}}}{1}
    {3}{{{\color{blue}3}}}{1}
    {4}{{{\color{blue}4}}}{1}
    {5}{{{\color{blue}5}}}{1}
    {6}{{{\color{blue}6}}}{1}
    {7}{{{\color{blue}7}}}{1}
    {8}{{{\color{blue}8}}}{1}
    {9}{{{\color{blue}9}}}{1}
    {"asset"}{{{\color{violet}"asset"}}}{7}
    {"id"}{{{\color{violet}"id"}}}{4}
    {"assetId"}{{{\color{violet}"assetId"}}}{9}
    {"assetType"}{{{\color{violet}"assetType"}}}{11}
    {"category"}{{{\color{violet}"category"}}}{10}
    {"classDid"}{{{\color{violet}"classDid"}}}{10}
    {"tBinding"}{{{\color{violet}"tBinding"}}}{10}
    {"standard"}{{{\color{violet}"standard"}}}{10}
    {"chain"}{{{\color{violet}"chain"}}}{7}
    {"contract"}{{{\color{violet}"contract"}}}{10}
    {"tokenId"}{{{\color{violet}"tokenId"}}}{9}
    {"identity"}{{{\color{violet}"identity"}}}{10}
    {"schemaVersion"}{{{\color{violet}"schemaVersion"}}}{15}
    {"identitySchema"}{{{\color{violet}"identitySchema"}}}{16}
    {"identifiers"}{{{\color{violet}"identifiers"}}}{13}
    {"identifierScheme"}{{{\color{violet}"identifierScheme"}}}{18}
    {"identifierValue"}{{{\color{violet}"identifierValue"}}}{17}
    {"jurisdiction"}{{{\color{violet}"jurisdiction"}}}{14}
    {"issuingAuthority"}{{{\color{violet}"issuingAuthority"}}}{18}
    {"taxonomies"}{{{\color{violet}"taxonomies"}}}{12}
    {"system"}{{{\color{violet}"system"}}}{8}
    {"code"}{{{\color{violet}"code"}}}{6}
    {"label"}{{{\color{violet}"label"}}}{7}
    {"spatialFootprint"}{{{\color{violet}"spatialFootprint"}}}{17}
    {"encoding"}{{{\color{violet}"encoding"}}}{10}
    {"geometry"}{{{\color{violet}"geometry"}}}{10}
    {"type"}{{{\color{violet}"type"}}}{6}
    {"coordinates"}{{{\color{violet}"coordinates"}}}{13}
    {"granularity"}{{{\color{violet}"granularity"}}}{13}
    {"documents"}{{{\color{violet}"documents"}}}{11}
    {"name"}{{{\color{violet}"name"}}}{6}
    {"hash"}{{{\color{violet}"hash"}}}{6}
    {"mediaType"}{{{\color{violet}"mediaType"}}}{11}
    {"issuedBy"}{{{\color{violet}"issuedBy"}}}{10}
    {"relations"}{{{\color{violet}"relations"}}}{11}
    {"relation"}{{{\color{violet}"relation"}}}{10}
    {"target"}{{{\color{violet}"target"}}}{8}
    {"attributes"}{{{\color{violet}"attributes"}}}{12}
    {"value"}{{{\color{violet}"value"}}}{7}
    {"unit"}{{{\color{violet}"unit"}}}{6}
    {"custom"}{{{\color{violet}"custom"}}}{8}
    {"localPolicyTag"}{{{\color{violet}"localPolicyTag"}}}{16}
    {"iotDeviceIds"}{{{\color{violet}"iotDeviceIds"}}}{13}
    {"sStatus"}{{{\color{violet}"sStatus"}}}{10}
    {"statusPurpose"}{{{\color{violet}"statusPurpose"}}}{15}
    {"statusListCredential"}{{{\color{violet}"statusListCredential"}}}{21}
    {"statusListIndex"}{{{\color{violet}"statusListIndex"}}}{17}
    {"sProof"}{{{\color{violet}"sProof"}}}{8}
    {"issuer"}{{{\color{violet}"issuer"}}}{8}
    {"issued"}{{{\color{violet}"issued"}}}{8}
    {"expires"}{{{\color{violet}"expires"}}}{9}
    {"sectionHash"}{{{\color{violet}"sectionHash"}}}{13}
    {"proofValue"}{{{\color{violet}"proofValue"}}}{12}
    {"compliance"}{{{\color{violet}"compliance"}}}{12}
    {"licenseId"}{{{\color{violet}"licenseId"}}}{11}
    {"sellableRegions"}{{{\color{violet}"sellableRegions"}}}{17}
    {"restrictions"}{{{\color{violet}"restrictions"}}}{13}
    {"effectiveFrom"}{{{\color{violet}"effectiveFrom"}}}{15}
    {"effectiveTo"}{{{\color{violet}"effectiveTo"}}}{13}
    {"regulatorDid"}{{{\color{violet}"regulatorDid"}}}{13}
    {"sStatus"}{{{\color{violet}"sStatus"}}}{10}
    {"statusPurpose"}{{{\color{violet}"statusPurpose"}}}{15}
    {"statusListCredential"}{{{\color{violet}"statusListCredential"}}}{21}
    {"statusListIndex"}{{{\color{violet}"statusListIndex"}}}{17}
    {"sProof"}{{{\color{violet}"sProof"}}}{8}
    {"issuer"}{{{\color{violet}"issuer"}}}{8}
    {"issued"}{{{\color{violet}"issued"}}}{8}
    {"expires"}{{{\color{violet}"expires"}}}{9}
    {"sectionHash"}{{{\color{violet}"sectionHash"}}}{13}
    {"proofPurpose"}{{{\color{violet}"proofPurpose"}}}{13}
    {"proofValue"}{{{\color{violet}"proofValue"}}}{12}
    {"custody"}{{{\color{violet}"custody"}}}{9}
    {"custodianDid"}{{{\color{violet}"custodianDid"}}}{13}
    {"location"}{{{\color{violet}"location"}}}{10}
    {"policy"}{{{\color{violet}"policy"}}}{8}
    {"auditCycleDays"}{{{\color{violet}"auditCycleDays"}}}{16}
    {"insurancePolicyRef"}{{{\color{violet}"insurancePolicyRef"}}}{19}
]
"id": "<did:example:subject>",
"asset": {
  "assetId": "<did:example:asset>",
  "assetType": "<string>"
  "category": "<string>"
  "classDid": "<did:example:class>",
  "tBinding": {
    "standard": "<string>(e.g.,'ERC-721')",
    "chain": "<string>(e.g.,'eip155:1')",
    "contract": "<string>",
    "tokenId": "<string>"
  }
},
"identity": {
  "schemaVersion": <number>,
  "identitySchema": "<URL>",
  "identifiers": [
    {
      "identifierScheme": "<string>",
      "identifierValue": "<string>",
      "jurisdiction": "<string>", 
      "issuingAuthority": "<string>"
    }
  ],
  //Attributes applicable to specific scenarios
  ...
  "sStatus": {...},
  "sProof": {...}
},
"compliance": {
  "licenseId": "<string>",
  "sellableRegions": ["<string>"],
  "restrictions": ["<string>"],
  "effectiveFrom": "<timestamp>",
  "effectiveTo": "<timestamp>",
  "regulatorDid": "<did:example:regulator>",
  "sStatus": {...},
  "sProof": {...}
},
"custody": {
  "custodianDid": "<did:example:custody>",
  "location": "<string>",
  "policy": "<string>",
  "auditCycleDays": <number>,
  "insurancePolicyRef": "<string>",
  "sStatus": {...},
  "sProof": {...}
}
\end{lstlisting}

When an RWA is deployed to a single blockchain, the tokenization process described above ensures that its digital representation is securely bound to a DID and a verifiable credential. 
However, in practice, RWAs often need to migrate across chains. 
A typical scenario resembles Web2 systems where an asset record, once canceled in one registry, must be reissued in another. 
Similarly, an RWA token may be burned on a chain $C_1$ and later reissued or transferred on $C_2$ to access new markets. 
Without interoperability, the tokenization procedure has to be repeated independently on each chain, leading to redundant authentication and increased operational overhead.  

To overcome this limitation, we next introduce a cross-chain authentication protocol that allows an RWA verified on one chain to be recognized on another without repeating the full tokenization workflow. 

\subsection{Cross-Chain Authentication} 
\label{sec:auth} 
Consider an RWA $a$ that has been tokenized on $C_1$, producing a credential $\mathsf{cred}$ anchored by a transaction $tx$ in block $B$. 
To enable recognition of $a$ on another chain $C_2$ without re-executing the full tokenization procedure, we employ SPV as defined in Section~\ref{subsec:SPV}.  

On $C_1$, the issuer embeds a commitment to $a$ within $tx$, which binds the asset identifier, the digest of the disclosed subset of $\mathsf{cred}$, the token binding, and auxiliary parameters such as an epoch and nonce. 
The transaction $tx$ becomes a leaf of the Merkle tree of $B$, with its membership authenticated by the Merkle root $\mu$ contained in the block header.  
To transfer authentication across chains, the issuer generates an SPV proof 
$\pi \leftarrow \text{SPV}.\mathsf{Prove}(tx, B)$. 
The proof $\pi$ = $(p,\mu)$ consists of the Merkle path $p$ linking $tx$ to the root and the Merkle root $\mu$ extracted from $B$. 
This compact proof, together with the serialized $tx$, is submitted to a verifier contract on $C_2$.  
The verifier then executes SPV.$\mathsf{Verify}(tx,\pi)$. 
It first checks that $\mu$ is consistent with a valid header of $C_1$ maintained under light-client rules. 
It then recomputes the path using $p$ to confirm that $tx$ is included under $\mu$. 
If the check passes, the verifier parses the embedded commitment from $tx$ and validates that it matches the disclosed fields of $\mathsf{cred}$, 
including issuer DID, revocation status, and jurisdictional restrictions. 
If all conditions hold, the credential $\mathsf{cred}$ is accepted on $C_2$ as if it were locally issued.  
Through this mechanism, authentication completed on $C_1$ can be securely reused on $C_2$. 

Unlike conventional approaches where each blockchain must independently repeat the full verification of decentralized identifiers and verifiable credentials, the SPV-based protocol requires such verification only once on the source chain. Subsequent chains merely validate the compact SPV proof to ensure that the authenticated transaction has indeed been recorded on the source ledger and that the embedded commitment is consistent with the disclosed credential fields. 

The detailed procedure of the SPV-based cross-chain authentication is summarized in Algorithm~\ref{alg:spv-auth}. 
\begin{algorithm}[!t]
\caption{SPV-based Cross-Chain Authentication}
\label{alg:spv-auth}
\begin{algorithmic}[1]
\State \textbf{Setup:} 
Asset $a$ is tokenized on $C_1$, producing a credential $\mathsf{cred}$ anchored by transaction $tx$ in block $B$.
\State \textbf{Commitment (on $C_1$):} 
Issuer embeds into $tx$ a commitment binding the asset identifier, digest of $\mathsf{cred}$, token binding, and auxiliary parameters. 
$tx$ is included in $B$, whose header contains the Merkle root $\mu$.
\State \textbf{Proof generation (off-chain):} 
Compute $\pi \leftarrow \text{SPV}.\allowbreak\mathsf{Prove}\allowbreak(tx,B)$, yielding $\pi$ = $(p,\mu)$, where $p$ is the Merkle path authenticating $tx$. 
Submit $(tx,\pi)$ to $C_2$. 
\State \textbf{Verification (on $C_2$):} 
Verifier executes $\text{SPV}.\allowbreak \mathsf{Verify}\allowbreak(tx,\allowbreak\pi)$. 
If valid, it parses the commitment in $tx$ and checks consistency with disclosed fields of $\mathsf{cred}$, 
including issuer DID, revocation status, and jurisdictional restrictions.
\State \textbf{Acceptance:} 
If all checks pass, $\mathsf{cred}$ is accepted on $C_2$ as if locally issued.
\end{algorithmic}
\end{algorithm}

\subsection{Efficient Cross-Chain Interactions}
In this part, we adopt HTLCs and off-chain channels to construct a cross-chain channel that supports partial settlement of RWAs without requiring channel closures. 
We first provide the definition of the cross-chain channel and then describe how it can be applied to the settlement of RWAs. 

%
%
\noindent {\bf Cross-Chain Channel.} A cross-chain channel~\cite{guo2023cross} aims to guarantee atomicity while significantly reducing on-chain interactions. 
It allows parties to perform frequent cross-chain payments off-chain and only settle on-chain when necessary. 
\begin{definition}[Cross-Chain Channel]\label{Def:CC}
The process of this channel can be represented by a tuple of polynomial-time algorithms  
$\Pi \overset{\text{def}}{=} (\mathsf{Open}, \allowbreak \mathsf{Update}, \allowbreak \mathsf{Lock}, \allowbreak \mathsf{Unlock}, \allowbreak \mathsf{Refund}, \allowbreak \mathsf{Close})$.  
    \begin{itemize}
        \item $\mathsf{Open}$:  
        Each party deposits assets into contracts on their respective chains, forming a bi-directional channel. 

        \item $\mathsf{Update}$:  
        During the channel lifetime, the two parties exchange signed off-chain states to update balances. 
        These updates do not require on-chain transactions, which improves efficiency. 

        \item  $\mathsf{Lock}$: The parties submit the negotiated interaction results to the blockchain and lock their respective assets. Hash-locked conditions $h(\rho)$ are placed on both chains with timeouts $T_1 > T_2$ to guarantee atomicity. 

        \item $\mathsf{Unlock}$:  
        If the receiver discloses the preimage $\rho$ within the time limit $T_2$, it can unlock the locked assets on one chain. 
        Once $\rho$ is revealed, the sender can use the same $\rho$ to unlock the corresponding assets on the other chain. 

        \item $\mathsf{Refund}$:  
        If the preimage $\rho$ is not disclosed before the timeout, the locked asset is returned on-chain to the original depositor’s account, 
        and the channel state is updated accordingly. 

        \item $\mathsf{Close}$:  
    When the channel is terminated, the remaining deposits locked in the channel are returned on-chain to the respective parties’ accounts, 
    thereby releasing the channel balance and concluding the channel lifecycle. 

    \end{itemize}
\end{definition}

The above definition shows that settlement of RWAs can be achieved within the channel lifecycle through the $\mathsf{Unlock}$ and $\mathsf{Refund}$ procedures, without requiring channel closures. 
This feature allows partial settlements to be executed while keeping the channel open, thereby reducing the cost of repeated channel establishment and improving the efficiency of cross-chain interactions. 

\noindent {\bf Cross-Chain Interactions.} 
First, the parties invoke $\Pi.\mathsf{Open}$ by deploying contracts $\xi^{\mathsf{1}}$ and $\xi^{\mathsf{2}}$ on $C_1$ and $C_2$, which establish the channel environment and record deposits.  
During the lifetime of the channel, they repeatedly apply $\Pi.\mathsf{Update}$ to exchange signed off-chain states describing tentative transfers of different subsets of RWAs. 
For example, a state $\sigma_j$ may specify that $\mathcal{B}$ pays $v_j$ units on $C_1$ in exchange for batch $S_j \subseteq \{a_1,\dots,a_m\}$. 
These updates remain entirely off-chain, enabling multiple negotiations without incurring blockchain overhead.  
Once the parties agree on a settlement, $\Pi.\mathsf{Lock}$ is executed: both sides upload the agreed final state to $\xi^{\mathsf{1}}$ and $\xi^{\mathsf{2}}$, where hash-locked conditions $h(\rho)$ with timeouts $T_1 > T_2$ are installed to ensure atomicity.  
If $\mathcal{B}$ reveals the preimage $\rho$ on $\xi^{\mathsf{2}}$ before $T_2$, $\Pi.\mathsf{Unlock}$ delivers the agreed union of RWAs $\bigcup_j S_j$ to $\mathcal{B}$ and allows $\mathcal{S}$ to redeem the corresponding aggregated payment on $\xi^{\mathsf{1}}$ before $T_1$.  
If instead $\mathcal{B}$ does not reveal $\rho$ in time, $\Pi.\mathsf{Refund}$ is triggered: $\xi^{\mathsf{2}}$ returns the RWAs to $\mathcal{S}$ at $T_2$, and $\xi^{\mathsf{1}}$ returns the funds to $\mathcal{B}$ after $T_1$.  
Finally, once the session concludes, $\Pi.\mathsf{Close}$ is called to release any remaining deposits and terminate the channel.  

Through this process, multiple RWAs can be traded within the channel via successive $\mathsf{Update}$ operations, while only a single round of on-chain $\mathsf{Unlock}$ or $\mathsf{Refund}$ is needed to settle the results. 
This reduces on-chain congestion and lowers the overall settlement costs. The detailed procedure of the cross-chain channel for multiple RWAs is summarized in Algorithm~\ref{alg:xrwa}. 
%

\begin{algorithm}[!t]
\caption{Efficient Cross-Chain Interactions}
\label{alg:xrwa}
\begin{algorithmic}[1]
\State \textbf{Participants:} Buyer $\mathcal{B}$ (funds on $C_1$), Seller $\mathcal{S}$ (RWAs $\{a_1,\dots,a_m\}$ on $C_2$).
\State \textbf{Contracts:} $\xi^{\mathsf{1}}$ on $C_1$, $\xi^{\mathsf{2}}$ on $C_2$.
\State \textbf{Parameters:} deposits $d_{\mathcal{B}}, d_{\mathcal{S}}$, preimage $\rho$, hash $h(\cdot)$, timeouts $T_1 > T_2$.
\vspace{1ex}

\State \textbf{Channel Opening (on-chain):}
\Statex \quad $\Pi.\mathsf{Open}$: initialize $\xi^{\mathsf{1}}$ and $\xi^{\mathsf{2}}$ and record deposits 
($\mathcal{B}\!\rightarrow\!\xi^{\mathsf{1}}$: $d_{\mathcal{B}}$, $\mathcal{S}\!\rightarrow\!\xi^{\mathsf{2}}$: $d_{\mathcal{S}}$). 
No hash-locked condition is installed at this stage.

\State \textbf{Off-chain Negotiation (loop):}
\Statex \quad $\Pi.\mathsf{Update}$: agree on batches $S_1,S_2,\dots,S_k \subseteq \{a_1,\dots,a_m\}$.
Each $S_j$ is embedded in a signed off-chain state $\sigma_j$; the aggregated state $\sigma$ encodes the intended settlement set $\bigcup_j S_j$ and the net payment.

\State \textbf{On-chain Locking (commit final state):}
\Statex \quad $\Pi.\mathsf{Lock}$: submit $\sigma$ to $\xi^{\mathsf{1}}$ and $\xi^{\mathsf{2}}$ and install hash-locked conditions 
$h(\rho)$ with timeouts $T_1 > T_2$ to ensure atomicity.

\State \textbf{Settlement (on-chain):}
\If{$\mathcal{B}$ reveals $\rho$ before $T_2$}
    \State $\Pi.\mathsf{Unlock}$ on $\xi^{\mathsf{2}}$: deliver $\bigcup_j S_j$ to $\mathcal{B}$.
    \State $\Pi.\mathsf{Unlock}$ on $\xi^{\mathsf{1}}$: release the aggregated payment to $\mathcal{S}$ before $T_1$.
\Else
    \State $\Pi.\mathsf{Refund}$ on $\xi^{\mathsf{2}}$: return $\{a_1,\dots,a_m\}$ to $\mathcal{S}$ at $T_2$.
    \State $\Pi.\mathsf{Refund}$ on $\xi^{\mathsf{1}}$: return $d_{\mathcal{B}}$ (and any unsettled funds) to $\mathcal{B}$ at $T_1$.
\EndIf

\State \textbf{Channel Continuation / Closure:}
The channel may remain open for further $\Pi.\mathsf{Update}$ rounds without reinstalling locks. 
Optionally, $\Pi.\mathsf{Close}$ releases remaining deposits and terminates the channel.
\end{algorithmic}
\end{algorithm}


\section{Security Analysis} \label{sec:security}
In this section, we provide the security analysis of xRWA. 
\begin{mytheorem}[Security of xRWA]
Assume that the underlying DID and VC infrastructure is secure, meaning that
no PPT adversary can cause a verifier to accept an identity or credential that
was not honestly issued by an authorized issuer, except with negligible
probability. Assume further that the SPV proof system is sound for
transaction inclusion, and that the cross-chain channel relies on a
collision-resistant hash function $h$ together with correctly enforced smart
contracts operating over immutable and consistent blockchains. Under these
assumptions, the xRWA protocol $\Pi_{\mathsf{xRWA}}$ achieves cross-chain
authentication correctness and cross-chain settlement atomicity, and therefore
satisfies RWA security as defined in Definition~\ref{def:rwa-security}.
\end{mytheorem}
\begin{proof}
Suppose there exists a PPT adversary $\mathcal{A}$ that breaks RWA security of $\Pi_{\mathsf{xRWA}}$ with non-negligible probability. By Definition~\ref{def:rwa-security}, this means that $\mathcal{A}$ either violates cross-chain authentication correctness or violates cross-chain settlement atomicity with non-negligible probability. We treat these two cases in turn and show that each yields an adversary that contradicts one of the assumed secure components. 

First, consider cross-chain authentication correctness. Assume that the event $\mathsf{AuthAcc}_{C_t}(a) \wedge \neg \mathsf{AuthRec}_{C_s}(a)$ occurs with non-negligible probability. Here $\mathsf{AuthAcc}_{C_t}(a)$ denotes that $C_t$ accepts $a$ via a cross-chain authentication claim from $C_s$, and $\mathsf{AuthRec}_{C_s}(a)$ denotes that $C_s$ contains a valid and confirmed authentication record for $a$. In xRWA, $C_t$ accepts such a claim only after verifying an SPV proof for a transaction on $C_s$ and checking that the embedded commitment matches a valid credential and issuer information.

We construct a simulator that uses $\mathcal{A}$ to break one of the assumed secure components. The simulator emulates honest parties and the blockchains for $\mathcal{A}$, and forwards its queries to the DID and VC system, or the SPV proof system, depending on the component being tested. Whenever $\mathcal{A}$ causes $\mathsf{AuthAcc}_{C_t}(a) \wedge \neg \mathsf{AuthRec}_{C_s}(a)$, the simulator inspects the transcript. If the SPV verifier accepts a proof for a transaction not contained in any valid block header of $C_s$, the simulator outputs an attack on SPV soundness. If the transaction exists but its commitment does not correspond to any honestly issued credential, or corresponds to a revoked or nonexistent credential, the simulator outputs an attack on the DID and VC infrastructure. If the credential was honest at issuance but later removed or altered while still being accepted by $C_t$, the simulator outputs an attack on ledger immutability. Hence, a successful violation of cross-chain authentication correctness by $\mathcal{A}$ leads to a successful attack on one of the assumed secure components with non-negligible probability, contradicting the assumptions. Therefore, the probability that $\mathcal{A}$ causes $\mathsf{AuthAcc}_{C_t}(a) \wedge \neg \mathsf{AuthRec}_{C_s}(a)$ must be negligible. 

Next, consider cross-chain settlement atomicity. Assume that the event $\mathsf{Com}_1 \oplus \mathsf{Com}_2$ occurs with non-negligible probability, where $\mathsf{Com}_i$ denotes that chain $C_i$ applies the settlement updates and $\oplus$ denotes exclusive or. In xRWA, all settlements are executed through the underlying cross-chain channel, which relies on the collision-resistant hash function $h$, correctly enforced smart contracts, and immutable blockchains, and which by assumption ensures atomic settlement. 

We construct a simulator that uses $\mathcal{A}$ to attack the cross-chain channel. The simulator runs $\mathcal{A}$ internally, forwards its messages to the channel challenger, and uses the challenger to emulate locking, settlement, and refund behavior. If $\mathcal{A}$ produces an execution in which exactly one chain commits the settlement, the simulator outputs this as an attack on channel atomicity, contradicting the assumed security of the channel. Consequently, the probability that $\mathcal{A}$ causes $\mathsf{Com}_1 \oplus \mathsf{Com}_2$ must be negligible.

Combining the two parts, any PPT adversary that breaks RWA security of $\Pi_{\mathsf{xRWA}}$ would enable an attack on one of the assumed secure components with non-negligible probability, contradicting their security. Hence, $\Pi_{\mathsf{xRWA}}$ satisfies cross-chain authentication correctness and cross-chain settlement atomicity, and therefore satisfies RWA security. 
\end{proof}

\section{Implementation and Performance Evaluation} \label{sec:per}
In this section, we present our concrete implementation of xRWA and test its performance. 

\subsection{Implementation}

The experiments are conducted on a workstation with an Intel Core i9-13980HX CPU and 32~GB RAM, running Windows~11 (24H2). 
Python~3.12.10 is used to implement and benchmark the VC routines and the SPV algorithm. 
The reported execution times are obtained from these Python implementations under multi-threaded settings. 
Smart-contract experiments are performed on Remix VM (Prague) using Solidity~0.8.30 to measure gas consumption for the HTLC and cross-chain channel workflows. 
The entire codebase consists of more than 2,000 lines, including the Python implementations for VC/SPV and the Solidity contracts for on-chain evaluation. Our code repository is publicly available at \href{https://github.com/ZhuHaoming2005/xRWA}{https://github.com/ZhuHaoming2005/xRWA}. 

\subsection{Performance Evaluation}
%
Table~\ref{tab:vc-size} presents the sizes (in KB) of VCs across different asset types. The covered categories include vehicles,  real estate (RE),  gold, art, bonds, funds, and intellectual property (IP). The sizes range from 6.95 KB to 8.05 KB, with an overall average of 7.27 KB. 
\begin{table}[!h]  \footnotesize
\centering
\caption{The size of VC across different RWA types.} 
\label{tab:vc-size}
\begin{threeparttable}  
\begin{tabularx}{0.9\linewidth}{|>{\centering\arraybackslash}m{3cm}|>{\centering\arraybackslash}m{3.65cm}|}
\hline    Type & Size (KB) \\  
\hline   Vehicle     & 7.11 \\
\hline   RE & 8.05 \\
\hline   Gold        & 6.95 \\
\hline   Art         & 7.38 \\
\hline   Bond        & 7.09 \\
\hline   Fund        & 7.27 \\
\hline   IP          & 7.10 \\

\hline  {\bf Average}     & \textbf{7.27} \\
\hline
\end{tabularx} 
\end{threeparttable}    
\end{table}

\begin{figure}[!h]
	\centering
	\includegraphics[width=0.7\textwidth]{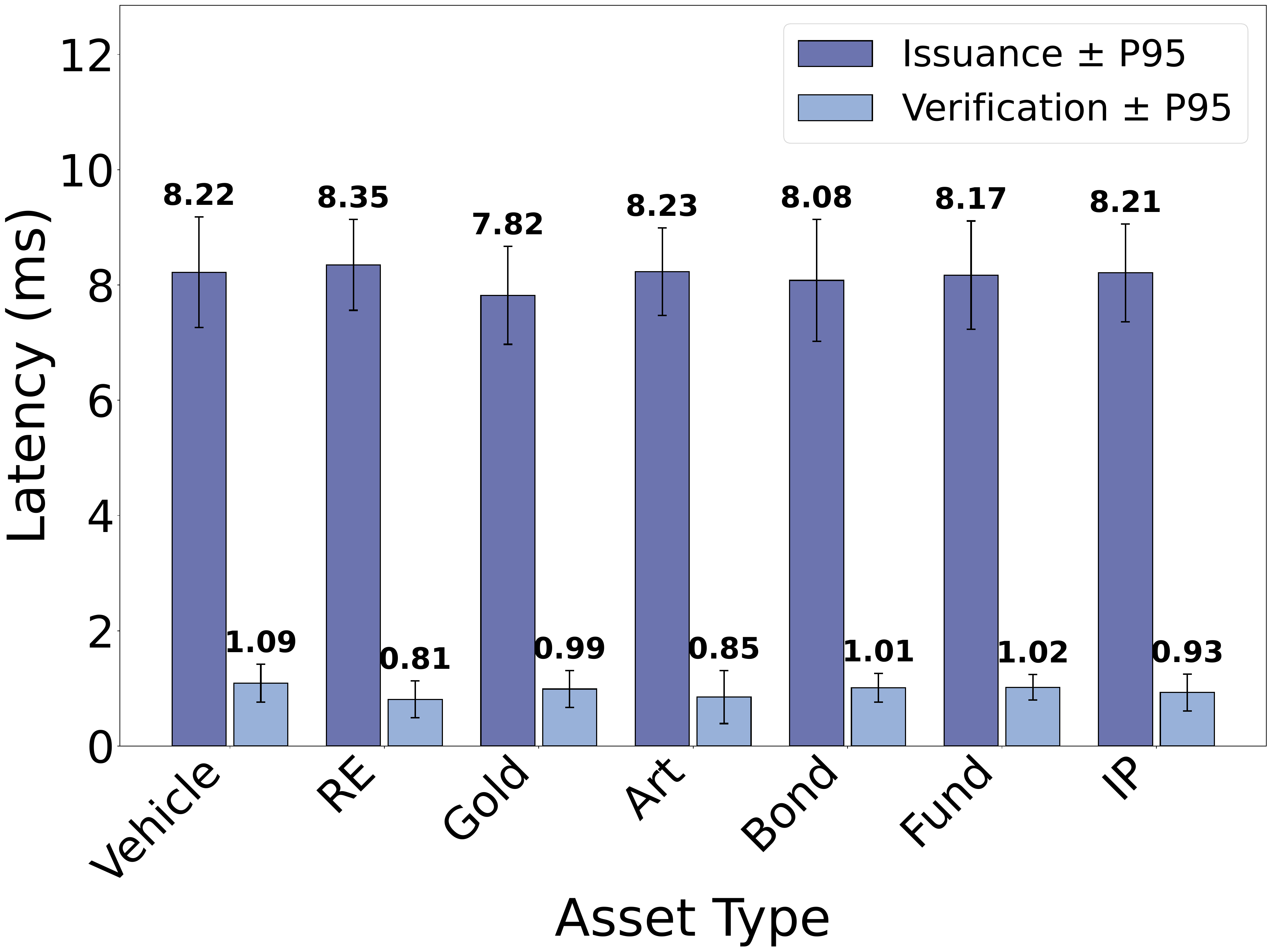}
	\caption{Latency in issuing and verifying  VCs.} 
\label{fig:vc-runtime}
\end{figure}

Figure~\ref{fig:vc-runtime} reports the performance of issuance and verification under a multi-threaded workload (500 credentials, 100 iterations, 8 worker threads). The average issuance latency is 8.16 ms per VC, with a P95 latency of 9.04 ms. Verification is faster, with an average latency of 0.96 ms per VC and a P95 of 1.27 ms. 
These results show that credential issuance requires on the order of several milliseconds per VC, while verification can be performed in less than one millisecond on average. The measured latencies indicate that both operations are lightweight and can be executed efficiently under multi-threaded workloads. 

\begin{figure}[!h]
	\centering
	\includegraphics[width=0.7\textwidth]{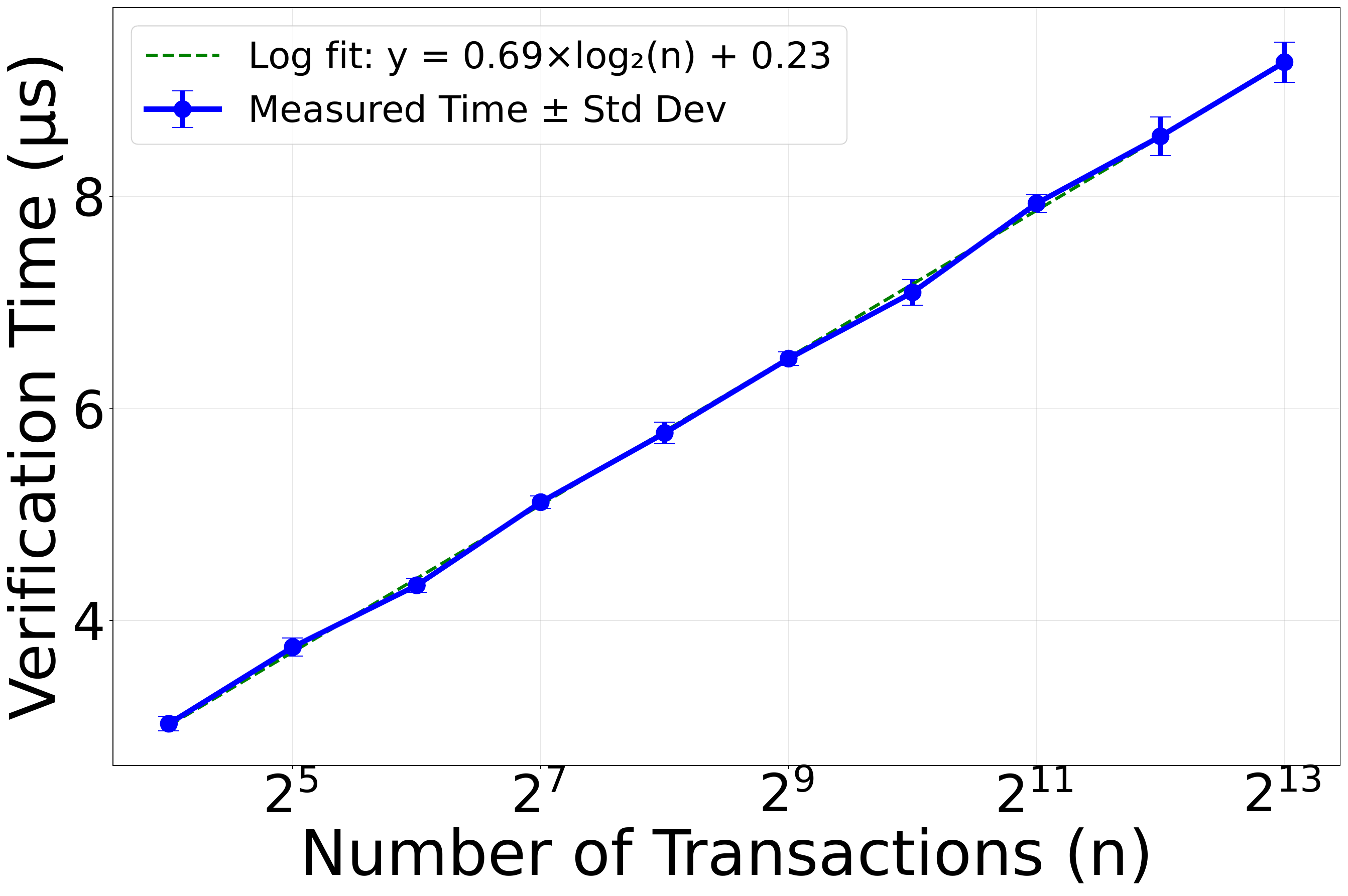}
	\caption{Verification time of SPV proofs under varying transaction counts.}
\label{fig:spv-performance}
\end{figure}

Figure~\ref{fig:spv-performance} reports the empirical validation of the $\mathcal{O}(\log n)$ complexity of SPV verification. 
The measured latency follows the fitted logarithmic model, where $n$ denotes the number of transactions per block.  
\begin{equation} \notag
t_{\text{SPV}}(n) \approx 0.69 \times \log_2(n) + 0.23 \ \text{$\mu$s}
\end{equation}
 
The experiments cover block sizes from $2^5$ to $2^{13}$ (32 to 8,192 transactions). 
Verification time increases from approximately $3.75$~$\mu$s at $n$ = 32 to $9.27$~$\mu$s at $n$ = 8,192, 
which is consistent with the expected logarithmic trend. 
This scalability shows that SPV can provide efficient proof validation even for large block sizes, 
making it a suitable basis for cross-chain RWA authentication.

For the cross-chain channel, handling $n$ cross-chain interactions requires only 
$917{,}253$ gas in total, since all updates are performed off-chain and only a small 
constant number of operations are executed on-chain. 
In contrast, the HTLC scheme consumes $465{,}426 \times n$ gas for $n$ interactions, 
as each interaction must execute both the lock and unlock operations on-chain, 
leading to a linear increase in cost. Moreover, compared with designs that close the channel after each cross-chain settlement, our workflow allows the channel to remain open so that it can either support the next cross-chain interaction or function as an intermediate relay, similar to a payment channel network, thereby avoiding the additional costs caused by repeatedly opening and closing the channel. 

In addition to the cost advantage, the channel mechanism also improves efficiency. 
The time overhead of cross-chain protocols is dominated by blockchain confirmation delays. 
Processing $n$ interactions with HTLC requires $4 \times n$ on-chain operations across two 
blockchains, and each operation incurs confirmation latency. 
By contrast, the cross-chain channel performs only four on-chain operations in total, 
regardless of $n$. 
Therefore, the channel mechanism provides both lower on-chain cost and higher efficiency compared with HTLC, particularly as the number of interactions increases.

\section{Conclusion} \label{sec:con}
This paper presents a cross-chain framework for RWAs that addresses identification, authentication, and interaction in a unified workflow. 
The design leverages DIDs and VCs for decentralization, applies an SPV-based protocol to prevent redundant verification, and introduces a cross-chain channel that enables partial settlement without closure. 
Experiments indicate that the framework reduces on-chain costs and improves efficiency in cross-chain settings. 
As future work, we explore potential security vulnerabilities in RWA projects and develop mitigation strategies to strengthen their reliability in practice.


\printcredits

\bibliographystyle{cas-model2-names}
\bibliography{references}

\end{document}